\documentclass[fleqn,usenatbib]{mnras}
\usepackage{newtxtext,newtxmath}

\usepackage[T1]{fontenc}
\usepackage{graphicx}	% Including figure files
\usepackage{amsmath}	% Advanced maths commands

\usepackage{xcolor}

\usepackage[outercaption]{sidecap}

\newcommand{\figref}[1]{Fig.~\ref{#1}}

\newcommand{\Figref}[1]{Fig.~\ref{#1}}
\newcommand{\Eqnref}[1]{Eq.~(\ref{#1})}

\newcommand{\be}{\begin{equation}}
\newcommand{\ee}{\end{equation}}
\newcommand{\eep}{\;\;.\end{equation}}
\newcommand{\eec}{\;\;,\end{equation}}
\newcommand{\bea}{\begin{eqnarray}}
\newcommand{\bel}[1]{\be\label{#1}}

\newcommand{\eea}{\end{eqnarray}}

\newcommand{\Bv}{\mathbf{B}}

\newcommand{\Rm}{\mbox{Rm}}

\newcommand{\B}{B}

\newcommand{\BAS}{\overline{B}}
\newcommand{\UAS}{\overline{U}}

\newcommand{\keltNb}{KELT-9b}
\newcommand{\UHJ}{UHJ}

\newcommand{\bs}[1]{\tilde{#1}}
\def\bm#1{\ensuremath{\mathchoice{\mbox{\boldmath$\displaystyle#1$}}
{\mbox{\boldmath$\textstyle#1$}}
{\mbox{\boldmath$\scriptstyle#1$}}
{\mbox{\boldmath$\scriptscriptstyle#1$}}}}
\usepackage{journals}
\definecolor{ashgrey}{rgb}{0.7, 0.75, 0.71}

\newcommand{\WD}[1]{\textcolor{black}{#1}}
\title{Magnetic induction processes in Hot Jupiters, application to KELT-9b}

\author[W. Dietrich et al.]{
Wieland Dietrich$^{1}$\thanks{E-mail: dietrichw@mps.mpg.de}, 
Sandeep Kumar$^{2,3}$,
Anna Julia Poser$^{4}$,
Martin French$^{4}$,
Nadine Nettelmann$^{5}$,
\newauthor{Ronald Redmer$^{4}$,
and Johannes Wicht$^{1}$}
\\
$^{1}$Max Planck Institute for Solar System Research, 37077 Goettingen, Germany\\
$^{2}$Center for Advanced Systems Understanding (CASUS), 02826 G\"orlitz, Germany\\
$^{3}$Helmholtz-Zentrum Dresden-Rossendorf (HZDR), 01328 Dresden, Germany\\
$^{4}$Universit\"at Rostock, Institut f\"ur Physik, 18051 Rostock, Germany\\
$^{5}$Deutsches Zentrum für Luft- und Raumfahrt,
Institut für Planetenforschung, 12489 Berlin, Germany
}

\date{Accepted XXX. Received YYY; in original form ZZZ}

\pubyear{2022}

\begin{document}
\label{firstpage}
\pagerange{\pageref{firstpage}--\pageref{lastpage}}
\maketitle
\begin{abstract} 
The small semi-major axes of Hot Jupiters lead to high atmospheric temperatures of up to several thousand Kelvin. Under these conditions, thermally ionised metals provide a rich source of charged particles and thus build up a sizeable electrical conductivity. Subsequent electromagnetic effects, such as the induction of electric currents, Ohmic heating, magnetic \WD{drag,} or the weakening of zonal winds have thus far been considered mainly in the framework of a linear, steady-state model of induction. For Hot Jupiters with an equilibrium temperature $T_{eq} > 1500$ K, the induction of atmospheric magnetic fields is a runaway process that can only be stopped by non-linear feedback. For example, the back-reaction of the magnetic field onto the flow via the Lorentz force or the occurrence of magnetic instabilities. Moreover, we discuss the possibility of self-excited atmospheric dynamos. Our results suggest that the induced atmospheric magnetic fields and electric currents become independent of the electrical conductivity and the internal field, but instead are limited by the planetary rotation rate and wind speed. As an explicit example, we characterise the induction process for the hottest exoplanet, KELT-9b by calculating the electrical conductivity along atmospheric $P-T$-profiles for the day- and \WD{nightside}. Despite the temperature varying between 3000 K and 4500 K, the resulting electrical conductivity attains an elevated value of roughly 1 S/m throughout the atmosphere. The induced magnetic fields are predominately horizontal and might reach up to a saturation field strength of 400 mT, exceeding the internal field by two orders of magnitude.
\end{abstract} 

  
\section{Introduction}
\label{intro}

Hot Jupiters (HJ) orbit their parent stars in very close proximity and are locked in synchronous rotation, which means that they always face the same side to the star \citep{Guillot1996,Showman2015}. The elevated electrical conductivity caused by thermal ionisation of metals and the fierce irradiation-driven winds induce strong electric currents. If these currents flow deeper into the \WD{atmosphere,}  the related Ohmic heating could explain the planetary radius inflation \citep{Batygin2010,Kumar2021}.
Some authors also suggest that Lorentz forces might
become strong enough to alter the atmospheric circulation and, thus, the brightness distribution \citep{Perna2010, Rogers2014}. Even a self-excited atmospheric dynamo that operates independently of the deep-seated, convective dynamo seems possible. \citet{Rogers2017} suggest that such dynamo action is promoted by the strong horizontal variation of the electrical conductivity caused by the large difference
between day- and nightside temperatures.

The interpretation of the observational data for Hot \WD{Jupiters} rely heavily on a reliable estimate of the electrical conductivity and the mechanisms of the induction process \citep{Batygin2010, Rogers2017}. So far, only a linear approximation of the induced electrical currents has been applied. However, this might be only justified in Hot Jupiters with equilibrium temperatures in the 1000-1500~K range, such as HD 209458b \citep{Kumar2021}.  

For hotter HJ, such as the group of Ultra Hot Jupiters (UHJ, $T_{eq} > 2200\,$K \citep{Parmentier2018}), the electrical conductivity, $\sigma_e$, might reach much higher values because the temperatures are sufficient to ionise more abundant metals, such as sodium, calcium or even iron. We therefore calculate the ionisation degree and electrical conductivity to characterise the induction process in a prominent UHJ, KELT-9b, the hottest planet detected so far. It orbits its host star, a main-sequence A0-type star, in only 1.48 days while receiving strong stellar irradiation~\citep{Gaudi2017}. It has a radius of $\approx 1.9 R_J$ and an age of 300--600 million years~\citep{Gaudi2017}.  It has been suggested that dayside and nightside temperatures are as high as 4600~K and 3040~K, respectively~\citep{Mansfield2020, Wong2020} . Another recent study reports dayside temperatures of up to 8500~K~ in the low pressure range of the atmosphere\citep{Fossati2020, Fossati2021}. Such high temperatures lead to significant ionisation and dissociation of atmospheric constituents~\citep{Hoeijmakers2019}, which
is confirmed by the observation of Fe$^{+}$ and Ti$^{+}$ \citep{Hoeijmakers2018, Hoeijmakers2019}.

In comparison with standard thermal evolution calculations, the observationally constrained radius of many Hot Jupiters appears inflated~\citep{Thorngren2018}, since their radii are larger than thermal evolution models suggest. This inflation seems particularly present for intermediately \WD{temperated} Hot Jupiter with equilibrium temperatures around 
between 1570~K \citep{Thorngren2018} and 1860~K \citep{Sarkis2020} depending on the atmosphere model used. Various studies investigate possible inflation mechanisms~\citep{Sarkis2020} and conclude that Ohmic heating is a promising candidate \citep{Batygin2010}. However, previous estimates of Ohmic heating were based on a steady and linear induction process that might not be readily applicable to hotter planets.

The importance of electromagnetic effects and the nature of the induction process can be characterised by estimating the ratio between the induction of magnetic fields and the dissipation. This is cast into the magnetic Reynolds number,
\begin{equation}
\label{Rm}
    \mbox{Rm} = \mu_0 \sigma_e U d\;\;,
\end{equation}
here $\mu_0$ is the vacuum permeability, $U$ is a typical flow velocity, $\sigma_e$ the electrical conductivity and $d$ a typical length scale. As long as Rm remains below one, the magnetic field induced in the outer part of the atmosphere will be smaller than the internal magnetic field originating from a convective, deep interior dynamo process. This allows estimating the induced field and the related electric current in a linear approach for an assumed internal field strength.

The linear approach has been used to predict magnetic effects caused by the observed zonal winds in the outer atmosphere of Jupiter and Saturn by \citet{Liu2008,Cao2017,Wicht2019}. 
In these planets, the electrical conductivity increases sharply with depth due to the growing ionisation degree of hydrogen until its transition into a metallic phase at around 1~Mbar. The electrical conductivity scale 
height is then the relevant length scale $d$ in equation \ref{Rm}. Adopting for example the conductivity model by~\cite{French2012} for Jupiter yields $d$ between $10^{-3} R_J$ and $10^{-2} R_J$, 
where $R_J$ is Jupiter's radius. 
As a consequence, Rm remains below one for the 
outer few percent in radius so that the linear \WD{approximation} for estimating electric currents and, thus, Ohmic dissipation can be applied. 

In contrast to Jupiter and caused by the irradiation driven ionisation of metals, for Hot Jupiters electromagnetic effects become important at much lower pressures, namely in the lower atmosphere around $1$ and $10^{-3}$ bar \citep{Kumar2021}. They are important when strong flows interact with a sizeable electrical conductivity. The permanent stellar irradiation will not reach  deeper than about 1 bar, where the infrared opacity reaches unity \citep{Iro2005}. Atmospheric flows driven by the irradiation gradients thus dominate above 1 bar.

In Hot Jupiters with a moderate equilibrium temperature of $1500\,\mathrm{K}$, such as HD 209458b with zero-albedo $T_{eq} = 1440$~K, the electrical conductivity reaches up to $\sigma_e \approx 10^{-3}\,\mathrm{S/m}$ \cite{Kumar2021}.  Assuming typical wind velocities of about $2\,\mathrm{km/s}$ yields $\mbox{Rm}\leq 1$. The linear approach for estimating the electric current is therefore still applicable and has been adopted by several authors~\citep{Batygin2010, Kumar2021}. 

For even hotter planets, such as the UHJ KELT-9b,
faster winds \citep{Fossati2021} and higher electrical conductivity (as we will show later)
will likely boost Rm to value much larger than one, $(\mbox{Rm}\gg 1)$. 
The linear approach becomes questionable and significant alteration of a deep dynamo field can be expected. The everlasting induction of atmospheric magnetic fields must be balanced by other processes. Even an independent self-excited atmospheric dynamo may  become possible, which could survive without the presence of a deep dynamo field.

In this paper, we discuss the induction of atmospheric magnetic fields in HJ and UHJ, its linear and non-linear processes and which amplitudes the induced fields and electrical currents can reach. This includes a general description of thermal ionisation of metals and the subsequent calculation of electrical conductivity. In a second step, we apply this to a specific planet by calculating the ionisation degree and the electrical conductivity in the atmosphere of the ultra-hot Jupiter KELT-9b with our previously published ionisation and transport model~\citep{Kumar2021}. Both quantities are calculated for several atmospheric profiles, including distinguished profiles for the day- and nightside. Lastly, we characterise the induction effects and give estimates for the induced magnetic field in the atmosphere.

Our paper is organized as follows. In Section.~\ref{sec:induction} we discuss the general induction of atmospheric magnetic fields and show how to characterise the induction process. In Section~\ref{elec_trans_sec} we apply the derived estimates and calculate explicitly the ionisation degree and the electrical conductivity along $p$-$T$ profiles of the atmosphere of  ultra-hot Jupiter KELT-9b. Conclusions are given in Section~\ref{concl}.

\section{Electromagnetic induction in Hot Jupiters}
\label{sec:induction}

\subsection{Induction of atmospheric magnetic fields}

 \begin{figure*}
     \centering
     \includegraphics[width=0.75\textwidth]{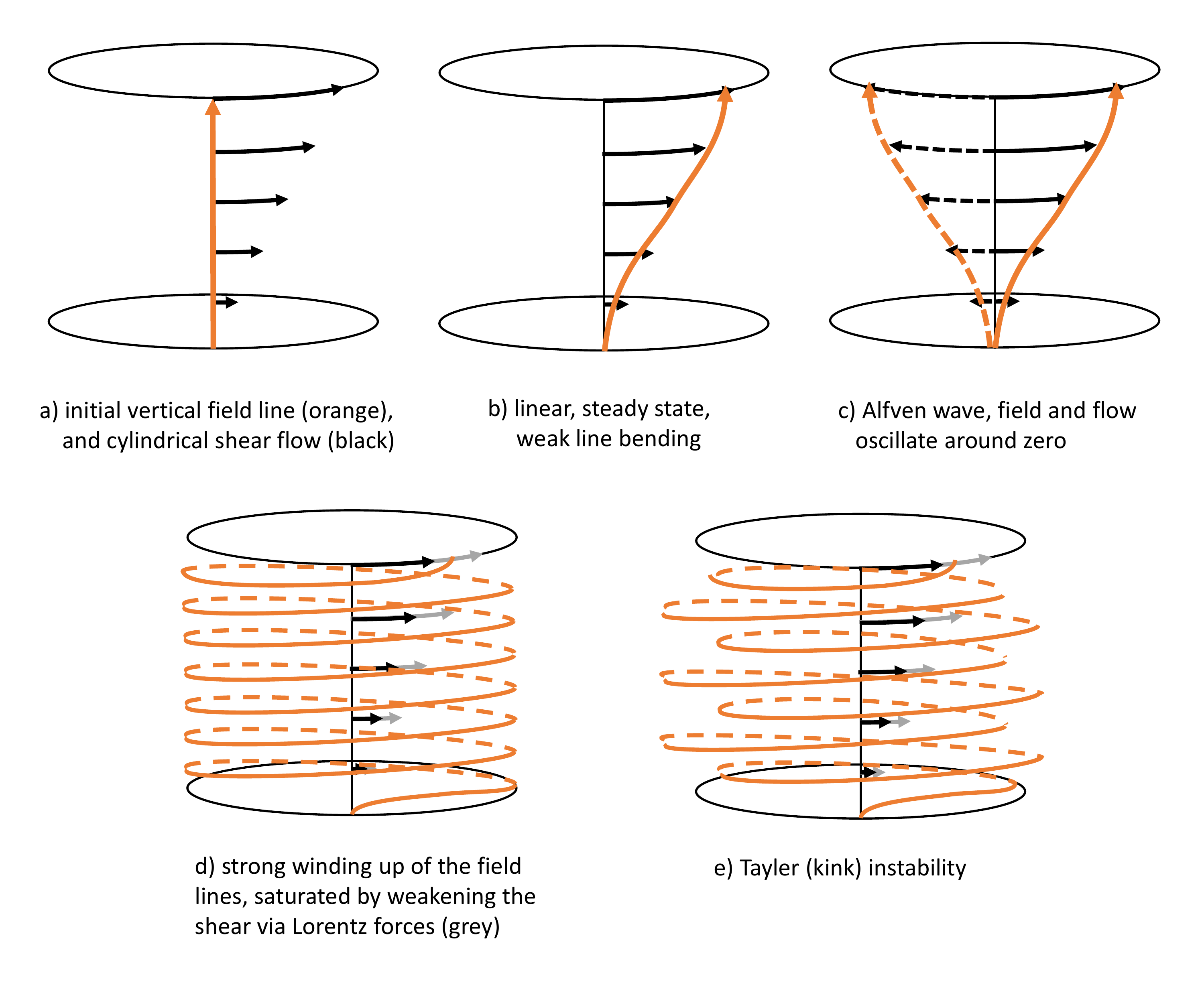}
\caption{Induction of azimuthal field by bending of field lines (orange) via atmospherical shear flows (black arrows). If the induction is stopped by efficient magnetic diffusion due to low electrical conductivity (panel b), the induced field remains weak compared to the background field and no Lorentz forces act on the flow. If the induction is rapid, the induced azimuthal field could quickly overpower the internal one and cause strong Lorentz forces. This could lead to the \WD{occurrence of Alfv\'en waves} flow and field oscillate (c). Or alternatively, that the Lorentz forces associated with the wound-up field ultimately weakens the shear flow (d). This \WD{tied-up} magnetic field tends to develop instabilities, such as the \WD{Tayler instability} with the kink mode (e). }
     \label{figindHJ}
\end{figure*}

The electric current, $\mu_0 \bm{j} = \bm{\nabla} \times \bm{B}$ in a moving conductor is given by Ohm's law:
\begin{equation}
    \bm{j} = \sigma_e \left( \bm{U} \times \bm{B}  + \bm{E} \right) \ ,
\end{equation}
where $\bm{U}$ is the flow, $\bm{B}$ is the magnetic and $\bm{E}$ the electrical field. Using the Maxwell equations this can be rewritten as the induction equation, which describes the evolution of magnetic fields by the competition of induction and diffusion:
\begin{equation}
    \partial_t \bm{B} = \WD{\bm{\nabla} \times  \left( \bm{U} \times \bm{B} \right) - \bm{\nabla} \times \left( \eta \bm{\nabla} \times \bm{B} \right) }\  ,
    \label{eqindgeneral}
\end{equation}
where $\eta=1/(\mu_0 \sigma_e)$ is the magnetic diffusivity.
The importance of the dynamo term relative to the diffusion term is estimated by the magnetic Reynolds number:
\begin{equation}
    \mbox{Rm} =  \frac{ \left\vert\bm{\nabla} \times  \left( \bm{U} \times \bm{B} \right)\right\vert} {\left\vert\bm{\nabla} \times \left( {\eta} \bm{\nabla} \times \bm{B} \right) \right\vert} \approx \frac{U }{\eta} \operatorname{min(d_\eta, d_U, d_B)} \ ,
\end{equation}
where $U$ is a typical flow velocity, $d_\eta, d_U, d_B$ characteristic length scales for electrical conductivity, the flow or the magnetic field, respectively. Estimating the relevant length scale of the induction or diffusion process is a key ingredient for $\mbox{Rm}$.

The induction becomes particularly simple if 
we assume that azimuthal zonal flows dominate the atmospheric dynamics 
while the radial flow component remains negligible. This 
seems a reasonable assumption for \WD{irradiation-driven}
flows in the stably stratified atmospheres of Hot and Ultra Hot Jupiters \citep{Showman2002}.

The induction process is then limited to creating azimuthal field (see Fig.~\ref{figindHJ}, a) via shear flows from a 
background field $B_{int}$ that is produced by the deep, internal dynamo process. If we furthermore assume that $\eta$ is constant, and that the background field, the induced azimuthal field, and the flow is predominantly axisymmetric (indicated by the overbar), the induction equation 
reduces to:
\begin{equation}
    \partial_t {\overline{B}_\phi} = - \overline{ \bm{B}}_{int} \cdot \bm{\nabla} \overline{\bm{U} } + \eta \Delta {\overline{B}_\phi} \;\; .
    \label{eqinBp}
\end{equation}

The stellar irradiation drives fast horizontal winds 
in the outer atmosphere of Hot Jupiters and UHJs. 
Since the winds likely remain confined to the low-opacity region
we expect a steep radial 
gradient in the wind profile with a typical scale height $d_U$ of   
\bel{eq:drU}
  d_U = \left|\frac{\UAS_\phi}{\partial \UAS_\phi / \partial r}\right|
\eep
Since numerical simulations indicate that the horizontal 
scales are typically large \citep{Showman2002}, we neglect the respective 
gradients in comparison to the radial one.

\subsection{Internal dynamo process}

Here we estimate the field strength of the internal dipole field, which is sheared by the atmospheric flows. 
As hydrogen becomes metallic at large 
pressures ($>1\,$Mbar) the electrical conductivity 
in the deep interior of gas planets reaches about $10^6\,\mathrm{S/m}$ \citep{French2012}.
We know that, Jupiter and Saturn host interior dynamos, and therefore it is likely, that Hot and Ultra-hot Jupiters also generate internal magnetic fields. 
The typical field strength in the dynamo regions
can be estimated based on scaling laws. 
Here we use such laws that rely on the available 
convective power which itself can be expressed 
in terms of the heat flux density $q_{int}$ out of the convective 
interior \citep{Christensen2009,Christensen2010}. The typical field strength for a self-sustained, convectively driven dynamo in gas giants scales like
\bel{eq:Brms}
  \B \propto f_{ohm}\, \rho_{c}^{1/6} \left( q_{int} H_\rho/H_T \right)^{1/3}
\eec
where $f_{ohm}$ is the ratio of Ohmic to total dissipation, $\rho_c$ is the
bulk density in the dynamo region, $q_{int}$ the heat flux at the radiative-convective boundary, $H_T$ the temperature scale height and $H_\rho$ the density scale height \citep{Christensen2009}.

We assume $f_{ohm}=1$, normalise this scaling relation with the well constrained values of Jupiter, and use the mean planetary density $\rho_c = 3 M_p / 4\pi  R_p^3$, where $R_p$ and $M_P$ are the radius and the mass of the Hot Jupiter under consideration:
\begin{equation}
    B_{int} = B_{int, J}\left( \frac{ R_J}{R_p}\right)^{1/2} \left( \frac{ M_p}{M_J}\right)^{1/6} \left(\frac{q_{int}}{q_{int, J}}\right)^{1/3} \ . 
    \label{eqBint}
\end{equation}
This is based on the assumption that the temperature and density scale \WD{heights} are similar. The surface field strength and interior heat flux of Jupiter is roughly $0.5\,\mathrm{mT}$ and $q_{int, J} = 5.4\,\mathrm{W/m^2}$.  

The internal heat flux density $q_{int}$ of the planet is difficult to access, since a variety of sources, such as secular cooling, gravitational contraction, tidal effects or Ohmic heating might contribute.

Classically, the amount of heat released from the radiative-convective boundary is primarily governed by the contraction and cooling during the thermal evolution of the planet. Thus larger and younger planets possess a larger luminosity. Thermal evolution models, e.g. \citet{Baraffe2003, Burrows2001} were used to derive scaling relations for the total luminosity as a function of age, radius and mass \citep{Burrows2001, Zaghoo2018}:
\begin{equation}
q_{int} = q_{int, J} \left( \frac{\tau_p}{\tau_J} \right)^{-1/3} \left( \frac{M_p}{M_J} \frac{R_p}{R_J} \right)^{2.64} \ ,
\label{eqthevo}
\end{equation}
where $\tau_p$ and $\tau_J = 4.5\,\mathrm{Gyrs}$ are the age of the planet and of Jupiter, respectively.

However, for Hot Jupiters, the heat budget seems more complex as the majority of them appear larger than thermal evolution models for non-irradiated planets would predict. This suggests additional heat sources that slow down secular cooling, and Ohmic dissipation is one of possible mechanisms \citep{Batygin2010, Sarkis2020}. This would indicate that \WD{Hot} Jupiters which receive more intense stellar radiation and are thus hotter tend to be more inflated by Ohmic dissipation since the higher \WD{electrical conductivity} allows for higher induced magnetic fields and currents, which transfer irradiation energy into the deeper interior through Ohmic dissipation. But if the atmospheric temperatures are too high, the Lorentz forces tend to suppress horizontal flows and thus Ohmic dissipation \citep{Menou2012}. 

If the atmosphere of Hot Jupiter is in thermal equilibrium with the incident flux, the internal heat flux and the one associated with  dissipating the irradiation must balance \citep{Thorngren2018, Sarkis2020}. Both studies found indeed statistical evidence for a relation between the internal heat flux due to extra heating and equilibrium temperature in the upper atmosphere with a pronounced maximum around $T_{eq}=1850\,\mathrm{K}$ \citep{Thorngren2019, Thorngren2020}:
\begin{equation}
 \WD{  T_{int} \approx  0.39 \, T_{eq} \exp{\left(-\frac{(\log(\sigma^*_{SB} T_{eq}^4) - 6.14)^2}{1.095}\right)}} \ ,
    \label{eqThorn}
\end{equation}
\WD{where $\sigma^*_{SB}$ is the Stefan-Boltzmann constant taken in units of $1/K^{4}$. The associated} internal heat flux can be calculated via $q_{int} = \sigma_{SB} T_{int}^4$. Note, that the exact mechanism of dissipating the irradiation is not undisputed \citep{Sarkis2020}.

\WD{Eqs.~(\ref{eqThorn})} and \WD{(\ref{eqthevo})} together with \WD{\Eqnref{eqBint}} provide a strategy to estimate the field strength of a convectively driven dynamo for a dedicated planet.

\subsection{Ionisation of metals and the electrical conductivity}
%%%%%%%%%%%%%%%%%%%%%%% Figure PT profiles
The main ingredient for assessing the electromagnetic induction effects,  quantified by $\mbox{Rm}$, is the electrical conductivity $\sigma_e$. Hot Jupiters reach atmospheric temperatures at which metals are partially or even fully ionised via thermal ionisation. We only consider the first degree of ionisation and therefore partial and full ionisation are to be understood with respect to the total abundance. Free electrons interact via collisions with ions, neutrals and other electrons to generate a macroscopic electrical resistivity. In the temperature range up to a few thousand of K, the relevant constituents that ionise are the alkali metals, such as lithium, potassium, sodium, rubidium, \WD{caesium}, and the common metals calcium and iron \citep{Lodders2003}. Hydrogen dominates the overall composition and acts mainly as scatterer for the electrons. It gets ionised only when temperatures exceed 5000 K. Here we use the solar system abundances for the atmospheric stoichiometry.
\begin{figure*}
\centering   
\includegraphics[width=\textwidth]{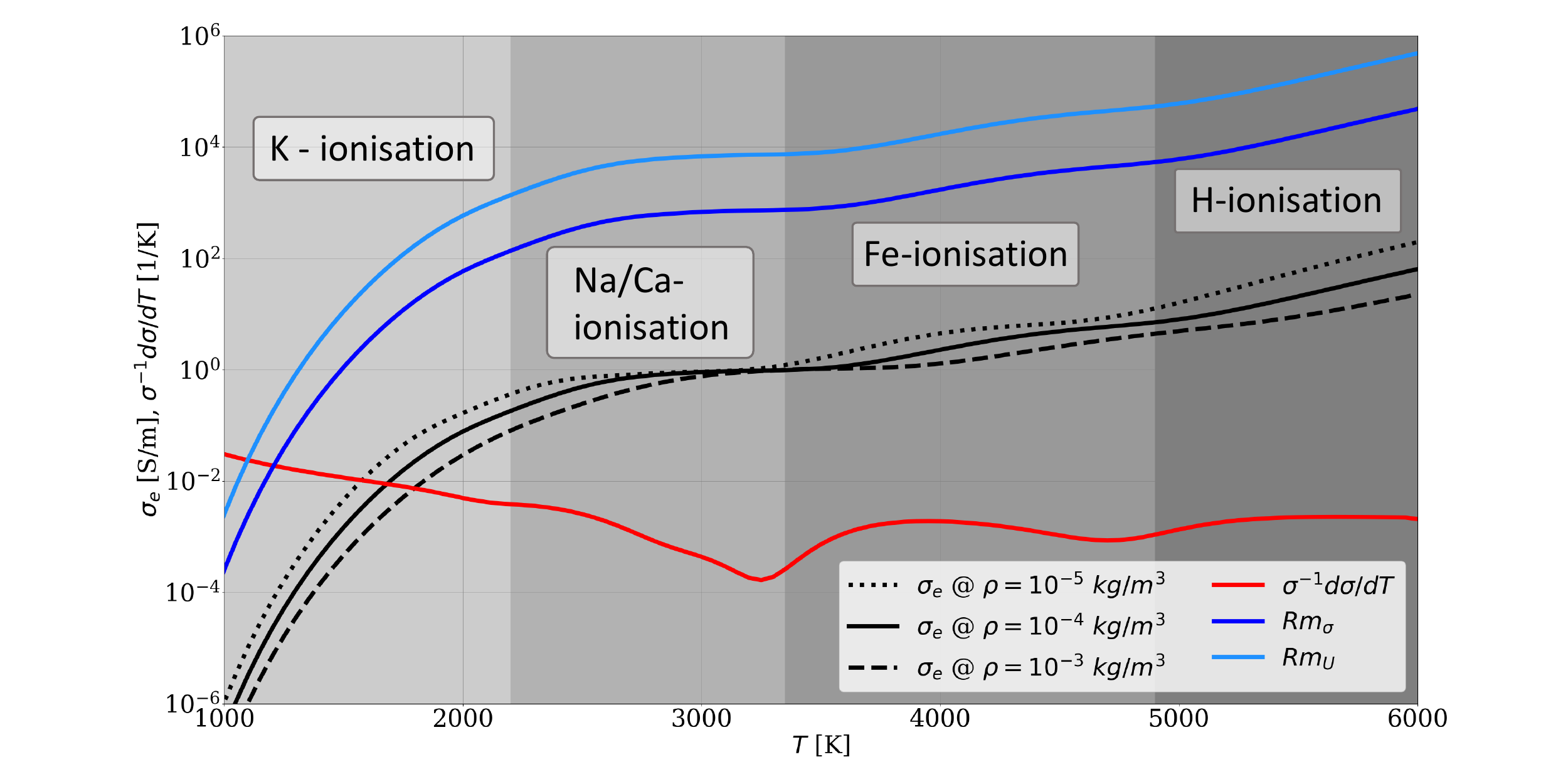}
\caption{Electrical conductivity due to ionisation of metals as a function of temperature along three isochores with $\rho=10^{-3}$ to $10^{-5}\,\mathrm{kg/m^3}$ (black solid, dashed and dotted line). The grey shaded areas indicate the main donors of electrons. The red profile shows the temperature variability of the electrical conductivity: $1/\sigma_e d\sigma_e / dT$. The blue profiles show two magnetic Reynolds numbers based on the solid conductivity profile, a flow velocity of $U=2\times 10^3\,\mathrm{m/s}$ \citep{Snellen2010}, and either $d_U = 3\times 10^6\,\mathrm{m}$ or $d_\eta =  3\times 10^5\,\mathrm{m}$ \citep{Wicht2019a} as length scale. Both curves cross unity around 1300 or 1500~\WD{K,} suggesting that the linear estimate of induction is not valid at temperatures \WD{higher} than that.}
\label{fig_conduc_overview}	                           
\end{figure*}
%%%%%%%%%%%%%%%%%%%%%%
For the calculation of the ionisation degree and electrical conductivity, we follow our model for a partially ionised plasma previously employed in \citet{Kumar2021}. There, mass-action laws are used to calculate the composition of the partially ionised plasma (PIP) \citep{Redmer1988, Redmer1999, Kuhlbrodt2000, Kuhlbrodt2005, Schottler2013}, from which the mass density $\varrho$ is derived for a given $p$-$T$ profile. Furthermore, the electrical conductivity is calculated from electron-ion and electron-neutral transport cross sections \citep{Martin2017}. The effect of electron-electron scattering on the transport properties is accounted for by introducing correction factors to the electron-ion contributions. The abundance of constituents considered in this work is the same (solar abundance) as we have considered in our PIP model \citep{Kumar2021}. 

For general illustration, we calculate $\sigma_e$ as function of temperature, but keep the density fixed to three different values between $\rho=10^{-3}$ and $10^{-5}\,\mathrm{kg/m^3}$ (Fig.~\ref{fig_conduc_overview}, solid, dashed and dotted black). The temperature axis can be interpreted as the radiative equilibrium temperature $T_{eq}$ or temperatures representing the day- or nightside. It can be seen that $\sigma_e$ increases with temperature as more and more electrons contribute to built up the electrical conductivity. However, the increase includes rather steep slopes and quite flat plateaus in between. These structures are caused by different species that are ionised subsequently as the temperature rises.

Given their high ionisation energy and small mass fraction, lithium, rubidium and caesium can safely be ignored in this consideration. Potassium on the other hand has the smallest ionisation energy and, thus, is the predominant source of electrons at low temperatures. It is strongly ionised already at temperatures below and up to 2000~K. At $T=2200$~K almost all of the free electrons stem from potassium. As an example, the Hot Jupiter HD 209485b has an equilibrium temperature of roughly 1500~K and thus a $\sigma_e \approx 10^{-3}\,\mathrm{S/m}$ \citep{Kumar2021}. Due to the partial ionisation of potassium, the temperature dependence around this value is relatively steep and small radial or azimuthal temperature variations might cause strong variations in $\sigma_e$.  

Between T=2300~K to 3500~K, the ionisation of calcium and sodium boost $\sigma_e$ to values between 0.1 and 1~S/m, where a plateau at an elevated electrical conductivity of 1~S/m is visible. Thus for Hot Jupiters which feature atmospheric temperatures in that range, $\sigma_e$ is rather independent and thus constant across the entire irradiated atmosphere. The width of this conductivity plateau is given by difference in ionisation energies between calcium/sodium and iron. Marking the high temperature end of the plateau (at T=3500~K) the next element providing more electrons is iron, which singly ionised at around $T=5000$~K. The last $\sigma_e$-boost is due to electrons from hydrogen which begin to ionise at T=5000K and dominate $\sigma_e$ at higher temperatures.

To show the parameter dependence, we display $\sigma_e$ along several isochores (dashed and dotted in \WD{Fig.~\ref{fig_conduc_overview}}) and calculate a temperature variability, $\sigma_e^{-1} d\sigma_e /dT$ (red profile ). It can be seen that a larger density smooths out the structures and decreases the magnitude of $\sigma_e$ due to a shift in the chemical ionisation equilibria according to \WD{le Chatelier’s} principle. 

The temperature dependence is most pronounced at the low-temperature end of the plot, where only a single species, potassium here, is partially ionised.
This leads to quite a strong temperature sensitivity of roughly $0.01\,\mathrm{1/K}$. At higher temperatures and in particular around $3300\,\mathrm{K}$ this drops down to $10^{-4}\,\mathrm{1/K}$ indicating that $\sigma_e$ is very insensitive to thermal variations. 

Moreover, we add estimates of $\mbox{Rm}$ in the figure (blue). Those are based on the actual value of $\sigma_e$, a ( conservative) flow amplitude of $U=2\times10^3\,\mathrm{m/s}$ and a length scale representing the conductivity scale height of $d_\sigma = 3 \times 10^5\,\mathrm{m}$ (dark blue) or the radial length scale of the shear flow $d_U = 0.02\,R_P \approx 3 \times 10^6\,\mathrm{m}$ (light blue). For the colder part, $\mbox{Rm}_\sigma$ might be more realistic, as $d_\sigma$ is smaller than $d_U$ and hence is the dominant diffusion length scale. Both curves are larger than unity from $T<1200$ and 1500~K suggesting that all HJ atmospheres exceeding these temperatures tend to host non-linear induction processes that are characterised by $\mbox{Rm} >> 1$. This already indicates that a better understanding of the atmospheric induction processes in atmospheres with high conductivity is strongly needed. 

\subsection{Linear regime, Rm$\le1$}
Previous approaches for estimating the magnetic induction 
in Hot Jupiters assumed that the growth is limited by
magnetic diffusion \citep[e.g][]{Batygin2010}.

As already discussed in the introduction and in \citet{Kumar2021}, 
for Hot Jupiters like HD 209458b, $\mbox{Rm}$ may actually remain below one,
because both, $\sigma_e$ and the characteristic length scale $d = d_\eta$ are rather small. The latter is the correct choice for the relevant diffusive length-scale to account for the strong variability of $\sigma_e$ \citep{Liu2008,Wicht2019a}.  This will likely apply to Hot Jupiters on the potassium-branch of \WD{Fig.~\ref{fig_conduc_overview}} or be the case in HJ atmospheres where T<1500\,K. For $\mbox{Rm}\le 1$, the dynamics establishes a quasi-stationary ($\partial_t B \approx 0$)
state where the magnetic dissipation balances the 
induction term in \WD{\Eqnref{eqinBp}}.
The locally induced field can then be estimated 
via
\begin{equation}
B_\phi = \mbox{Rm} \,  B_{int} \ .
\end{equation}
 This implies that 
the maximum amplitude of the induced field where this approximation is still valid is the internal field strength. Thus the radial field lines will be only slightly bent in the direction of the shear as indicated in \WD{Fig.~\ref{figindHJ},~b}.

Another consequence of the quasi-stationarity is that the electric currents obey a simplified Ohm's law where the gradient of the electric potential can be neglected 
\begin{equation}
    \bm{j} = \sigma_e \left( \bm{U} \times \bm{B}_{int} \right) \;\;,
\end{equation}
and can thus be calculated for a given flow and background field if we assume that $j = \sigma_e \overline{U} \, \overline{B}_\phi$ \citep{Wicht2019a}. It is important to \WD{stress} that the local currents and induced magnetic fields depend on the electrical conductivity, the internal field strength and the zonal wind speed. In this scenario, the induced field \WD{cannot} reach sufficient strength to modify or abate the flow via Lorentz forces (\WD{Fig.~\ref{figindHJ}, b)}.

\subsection{Non-linear induction: winding up the field}
If $\sigma_e$ is larger or the atmospheric winds more energetic, the induction equation (\Eqnref{eqinBp}) shows that the zonal flow shear will very efficiently induce azimuthal magnetic field from winding up the radial background field. If the rate of field induction cannot be balanced by the diffusion alone, a steady linear model as discussed before can not be applied anymore. As long as the radial field is provided from the deep, internal dynamo, the winding up of magnetic field lines around the \WD{planet} continues until a non-linear effect stops this process. This by itself does not work as a self-sustained dynamo, for which the atmospheric dynamics would itself replenish the radial field and thus become independent of the internal field. We discuss the requirements of the self-excited dynamo in sec.~\ref{secdynamo}.

The rapid induction process is represented by magnetic Reynolds number that strongly exceeds unity. Thus the diffusive term can be ignored and  
the induction equation (\Eqnref{eqinBp}) simplifies to
\bel{eq:Ind0}
 \frac{\partial \BAS_\phi}{\partial t} =
     \frac{\UAS}{d_U}\;\BAS_{int}
\eep

The atmospheric shear flow efficiently produces an azimuthal field component by shearing the radial component of the background field. It effectively winds up the field lines in azimuthal direction 
as illustrated in Fig.~\ref{figindHJ}, b and d. 
The induced field $\BAS_\phi$ thus 
increases linearly with time until some critical strength is reached  where the process is modified.

The related electrical current must then be calculated from the curl of the azimuthal field $B_\phi$. If, the induced field remain axisymmetric this is given by:
\begin{equation}
    j = \left|\frac{1}{\mu_0} \bm{\nabla} \times {\bm{B}}\right|\approx \frac{\overline{B}_\phi}{\mu_0 d_U} \label{eq:gencurrent}\ , 
\end{equation}
where we assumed that the electrical current is dominated by the $\partial_r(r\overline{B}_\phi)$-term of the
latitudinal component. 

\subsubsection*{Alfv\'en Waves}
The interaction between the current of induced azimuthal field and internal field defines a Lorentz force $\bm{j} \times \bm{B}$, that accelerates $\overline{U}_\phi$ in the opposite direction. Using the axisymmetric approximation of the electrical current \WD{in \Eqnref{eq:gencurrent} leads to}
\bel{eq:LF}
\rho \frac{\partial \overline{U}_\phi}{\partial t} = - \frac{\overline{B}_\phi \overline{B}_{int}}{\mu_0  d_U} 
\eep
The action of the Lorentz force thereby reduces
the radial shear and thus the growth rate of $\BAS_\phi$. 
\Eqnref{eq:Ind0} and \WD{(\ref{eq:LF})} define a wave equation for the azimuthal flow:
\begin{equation}
   \frac{\partial^2 \overline{U}_\phi}{\partial t^2}  = - \frac{1}{ \mu_0 \rho} \frac{B_{int}^2}{d_U^2} \overline{U}_\phi \ .
\end{equation}
This describes an \WD{Alfv\`en} wave (see \WD{Fig.~\ref{figindHJ}, c}) that travels along the field lines of the internal field and has a frequency 
\bel{eq:VA}
   \omega_A = \frac{\overline{B}_{int}}{d_U\,\left(\mu_0\rho\right)^{1/2}}
\eep
There is ample time for the winding action to take place before the internal field changes. We expect Lorentz forces and Alfv\`en waves to play an important role for atmospheric dynamics.
When assuming that $\UAS_\phi$ represents the maximum flow amplitude, the maximum field amplitude of the wave is 
\bel{eq:Bphi1}
 \BAS_\phi = \left(\mu_0 \rho\right)^{1/2}\; \UAS_\phi
\eep

Since this estimate neglects diffusive effects and assumes that $\UAS_\phi$ represents a pure Alfv\'en wave, it can only serve as an upper bound. However, Alfv\'en waves are well studied MHD phenomena that might play an important role in the dynamics of Hot Jupiter atmospheres. In fact, the study of \citet{Rogers2017b} performing MHD simulations of the irradiated atmosphere of the Hot Jupiter HAT-P-7b showed that temporal variability of the winds stems from magnetic effects. This is in line with the observed variability of the hot spot offset for that specific planet. In general, the \WD{hot} spot offsets are typically prograde and not very time-dependent \citep{Parmentier2018}.

%%%%%%%%%-----RADIAL FORCE BALANCE ----------
\subsubsection*{Radial Force Balance}
The derivation of the Alfv\'en-wave above was based on neglecting the Coriolis force. A balance between Coriolis and Lorentz \WD{forces} is typically expected in dynamo theory. We thus study the non-azimuthal component of this force balance: 
\begin{equation}
    \bs{\rho} \bm{\Omega} \times \bm{U} \approx \frac{1}{\mu_0} \bm{j} \times \bm{B} \ .
\end{equation}
Choosing the axisymmetric radial component of this balance involves the zonal flow $\overline{U}$, the zonal magnetic field $\overline{B}_\phi$ and \WD{\Eqnref{eq:gencurrent}} yields 
\begin{equation}
    {\rho} \Omega \overline{U}_\phi \approx \frac{\overline{B}_\phi^2}{\mu_0 d_U}\;\;.
\end{equation}
We find for the azimuthal field a saturation value of:
\begin{equation}
    B_\phi \approx \left( \mu_0 {\rho} d_U \Omega \overline{U}\right)^{1/2}
    \label{eqforcebalance}
\end{equation}
Physically, this could be interpreted such that meridional circulation cells redistribute zonal angular momentum in such a way that the radial shear is suppressed (Fig.~\ref{figindHJ}, d).
It is then the magnetic tension that saturates the initial shear flow (grey arrows) at a weaker final amplitude (black).

\subsubsection*{Tayler-instability}
There are different magnetic instabilities that 
can limit the growth of $\BAS_\phi$. 
For a introduction we refer to \citet{Spruit1999}. 
In a stably stratified layer like the outer atmosphere
of Hot Jupiters, the so-called Tayler-instability 
is likely to set in first. This instability refers to the unconstrained growth of non-axisymmetric magnetic field contributions when the axisymmetric part is sufficiently wound-up (see \WD{Fig.}~\ref{figindHJ}, e). A condition is that the atmospheric environment is sufficiently stratified and that Alfv\'en waves are slower than the rotation frequency 
\bel{eq:TC}
 N \gg \Omega \gg \omega_A
\eec
where $N$ is the Brunt-V\"ais\"al\"a frequency that 
characterises the degree of stratification. $N$ can be calculated from observable quantities if the atmosphere is to first order isothermal:
\begin{equation}
    N_T = \frac{g_P}{\sqrt{c_p T_{eq}}} \ ,
\end{equation}
here $g_P$ is the gravity and $T_{eq}$ the equilibrium temperature of the planet. $N$ typically exceeds the rotation rate $\Omega_P$ by a factor of 30 to 300. 
The Alfv\'en frequency of the wound-up 
azimuthal field is given by:
\bel{eq:AV}
 \omega_A = \frac{\BAS_\phi}{R_P \left(\mu_0\rho\right)^{1/2}}
\eep
The relation \WD{\Eqnref{eq:TC}} should be fulfilled for most HJ, when the induced field are not exceedingly large.

%--------------CHECK THIS CONDITION FOR ALL HJ --------

The most unstable azimuthal wave number is $m=1$ and
thus the instability assumes the form of predominantly horizontal
displacements on a small radial scale 
(see \WD{Fig}.~\ref{figindHJ}, e). This mode is also called the kink instability.
The details of the analytical description are discussed in the appendix. The critical field strength for
the onset of the \WD{Tayler instability} is given by \citep{Spruit2002}: 
\bel{eq:BphiTayler}
  \B_{\phi} = \Omega R \left(\mu_0\rho\right)^{1/2}\;
   \left(\frac{N}{\Omega}\right)^{1/2} \left(\frac{\eta}    
   {R^2\Omega}\right)^{1/4}
\eec
which saturates at a field strength of 
\bel{eq:Binst}
   \B_\phi = \left(\mu_0 \rho\right)^{1/2}\;
                \frac{\Omega}{N}\frac{R\,\UAS}{d_r} 
\eep
A characteristic of the instability mechanism is the 
displacement of the original wound-up field.
We suggest that this displacement 
seeks to increase dissipation until this 
balances the growth without \WD{changing the} field strength. 
The \WD{estimate}, i.e. \Eqnref{eq:BphiTayler}, then still holds, but
the magnetic field structure changes in a 
way illustrated in \figref{figindHJ}, e. 
This heuristic view seems to be supported by 
a numerical simulation where axisymmetric zonal field
and Tayler-instability field assume a similar amplitude
\citep{Zahn2007}.

\subsection{Ohmic heating}
For the linear case of limited, weak atmospheric induction characterised by $\mbox{Rm}\le 1$, the induced electrical currents are given by \WD{$j_{lin}=\sigma_e \overline{U}_\phi\overline{B}_{int}$} and the Ohmic dissipation is: 
\bel{eq:Ohmic:lin}
   P_{lin} = \int_{V} \sigma_e \overline{U}^2_\phi B_{int}^2 dV \ .
\ee
In this limit, the Ohmic power scales with electrical conductivity and the square of the internal field strength.

If $Rm>1$, the non-linear nature of the induction process makes it necessary to find the currents via $\bm{j} = \mu_0^{-1} (\bm{\nabla} \times \bm{B})$. However, if the induced field remains axisymmetric, the strongest gradient remains the radial derivative of the azimuthal field and thus $j = B_\phi/(\mu_0 d_U)$ (see \Eqnref{eq:gencurrent}). Assuming that the non-axisymmetric Tayler-instability preferentially develops an $m=1$-azimuthal structure, this approximation still holds. This then yields an Ohmic power of
\bel{eq:Ohmic:nl}
   P_{nl} = \frac{1}{\mu_0^2}\int_{V} \frac{B_\phi^2}{\sigma_e d^2_U} dV \ .
\ee
Since the induced field $B_\phi$ is independent of $\sigma_e$, the Ohmic dissipation for the non-linear case shrinks rather than grows with higher $\sigma_e$.

Most of the dissipation happens in the atmospheric region, where the currents are induced initially. One can thus expect that just a small fraction of the currents \WD{connect} down to the convective interior and could potentially interfere with the secular cooling \citep{Batygin2010}.

Consequently, Ohmic heating was proposed as one of the mechanisms that could explain the radius anomaly of Hot Jupiters \citep{Batygin2010,Thorngren2018, Sarkis2020}. Radial currents would flow to deeper regions below the radiative/convective boundary where the related Ohmic heating 
\bel{eq:Ohmic}
   P = \int_{V_i} dV\;\frac{j^2}{\sigma_i}
\ee
could explain the inflation. Here $V_i$ denotes an integration
over this deeper region with electrical conductivity $\sigma_i$.
The currents are induced in the wind region and decay in the 
deeper region that has been denoted 'leakage region' by \citet{Kumar2021}. Relevant for the amount of deeper heating are 
i) the current strength in the induction layer above, ii) the deeper electrical conductivity, 
but also iii) 
the electric current pattern at the transition to the deeper region, which determines how deep 
the currents would penetrate. 
Here we assume that the pattern is 
dominated by the action of large scale zonal winds
on a large scale internal field and therefore generally very similar.

\subsection{\WD{Self-sustained} Atmospheric Dynamo}
\label{secdynamo}
In a self-excited dynamo, a small seed field will be amplified by the atmospheric flows until the associated Lorentz forces are strong enough to sufficiently modify these flows. Here we analyse whether the prerequisites for self-sustained dynamo action are met and evaluate the possibility of atmospheric dynamo action. Modelling this complex \WD{phenomenon} requires dedicated MHD numerical simulations, that are beyond the scope of the present study.

A dynamo can only be maintained when various conditions on the flow amplitude, \WD{direction,} and complexity are met. 
A necessary condition is that Rm is sufficiently larger than 
one. This is certainly true for the azimuthal (toroidal) field 
generation discussed above. 
However, for an independent atmospheric dynamo that could 
operate even without a background field, the 
process also has to generate radial (poloidal) field from
the azimuthal (toroidal) field. 
This generally requires radial flows fast enough so 
that the respective magnetic 
Reynolds number $\mbox{Rm}_r = \mu_0 \sigma_e u_r d_r$ to exceed one. For this condition to be fulfilled, it seems sufficient that $u_r  \sigma_e > 1\, \mathrm{S/s}$.

In the study by \citet{Showman2002}, the non-zonal part of the irradiation driven winds in numerical models for Pegasi 51 b amounts to 20 m/s. More recently, the analysis of the vertical mixing rates in a broad suite of general circulation models (GCM) suggested radial flows on the order of 2 and 20 m/s \citep{Komacek2019}.
This already suggests that independent atmospheric dynamos are possible where the dynamics of the irradiated atmosphere could generate magnetic fields even when there is no deep-seated dynamo. A more definitive answer to this question would require 3D simulations of the induction processes in the dynamic atmosphere. 

 A relevant alternative process to induce radial field was suggested by \citet{Busse1992} and \citet{Rogers2017} and is driven by the lateral variations in electrical conductivity. As along as these variations are relatively modest, this process is too inefficient.

Furthermore, \citet{Spruit2002} envisioned self-sustained dynamo action for stars where the field generated by the Tayler-instability would replace the interior field and replenish the reservoir of the poloidal field component. An alternative mechanism turning the atmospheric induction into a \WD{self-consistent} dynamo is the strong horizontal variation of electrical conductivity. This was theoretically predicted by \citet{Busse1992} and numerically investigated by \citet{Rogers2017}.  

Note, that a self-sustained dynamo might generate a magnetic field and thus dissipate magnetic energy at much smaller length scales. This would in fact drastically increase the total Ohmic power.

\section{Application to Hot Jupiters}
\WD{
Before we apply the ionisation model to atmospheric $P$-$T$-profiles of two dedicated Hot Jupiters, we give a general, order of magnitude, assessment of \mbox{Rm}. According to \Eqnref{Rm} this requires a flow speed, a length scale and the electrical conductivity. The zonal atmospheric winds driven by irradiation gradients reach amplitudes of several km/s, hence even exceeding the angular velocity at the planetary surface due to its solid body rotation. A conservative estimate for the zonal flow is thus $\UAS_\phi \ge \Omega R$. For the length scale, we use the depth of the winds $d \approx 0.02 R$. Note, that for colder HJ, $\sigma_e$ might be very temperature dependent and hence the radial conductivity scale height could the relevant (dissipative) length scale. }

\begin{figure*}
\centering   
\includegraphics[width=\textwidth]{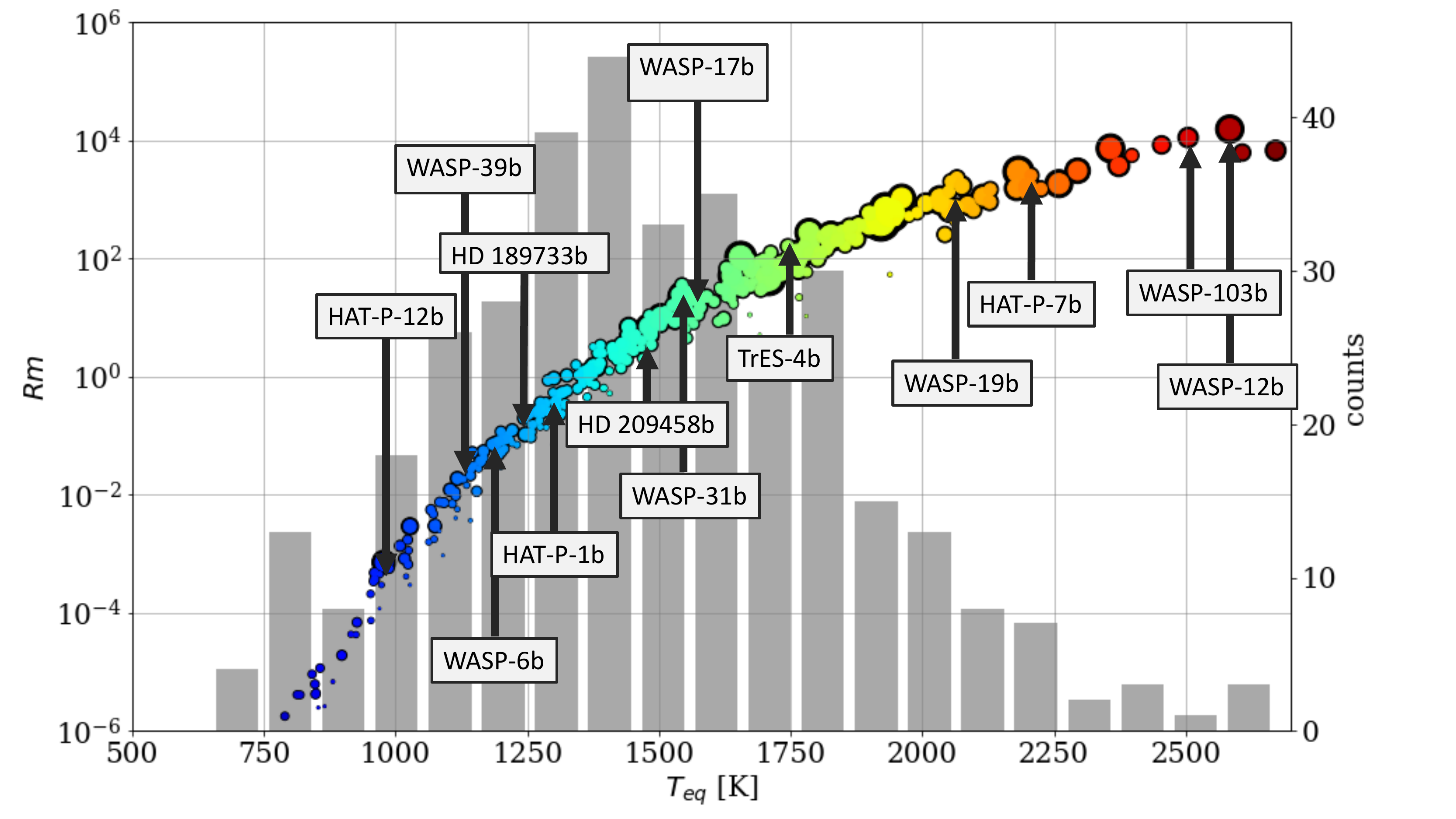}
\caption{\WD{Estimate of \mbox{Rm} for Hot Jupiters (HJ) as a function of $T_{eq}$ for a density of $10^{-4}~\mathrm{kg/m^3}$. The colour refers to the equilibrium temperature, the size of the dots to the radius. The histogram shows the population of the selected group of HJs. A few popular planets are marked. For $T_{eq} > 1300\,$K, Rm exceeds unity. Thus, roughly the colder half of HJs host strongly non-linear atmospheric induction processes.}
}
\label{fig_Rm_overview}	                           
\end{figure*}

\WD{The most influential quantity is the electrical conductivity, $\sigma_e$, which can vary by orders of magnitude (see Fig~\ref{fig_conduc_overview}). For a quick characterisation of HJ, we calculate $\sigma_e$ for a fixed density and as a function the respective atmospheric equilibrium temperature $T_{eq}$ . This gives a planet-specific estimate of \mbox{Rm} by:  
\begin{equation}
    Rm_p \approx 0.02 \mu_0 \, \sigma_e(T_{eq}) \,  \Omega \, R^2 \ .
\end{equation}
Note, that this \mbox{Rm} estimate is calculated only from (apart from $\sigma_e$) on observable quantities, such as the orbital rotation rate or the planetary radius. \Figref{fig_Rm_overview} shows the so derived \mbox{Rm}-values for set of ca. 350 HJ (masses between 0.1 and 10 Jupiter masses, semi-major axis smaller than 0.1~AU and radii between 0.5 and 2.1 Jupiter radii) as a function of their $T_{eq}$. It can be seen, that \mbox{Rm} exceeds unity at $T_{eq} \approx 1400$~K. For all planets below that temperature the linear induction model for which $Rm<1$, can readily be applied. For all planets with temperatures significantly exceeding this threshold the atmospheric induction of electromagnetic currents is a run-away process that can only be saturated by non-linear feedback. 
}

Here we study two examples, HD 209458b and KELT-9b\WD{. Both are} well-characterised Hot or Ultra-Hot Jupiters, but with very different characteristics of electromagnetic induction in the irradiated atmosphere. \WD{Whereas HD 209458b is an example of $\mbox{Rm} \approx 1$, KELT-9b is the hottest HJ observed so far with $\mbox{Rm} >> 1$.  } 

\WD{As discussed in sec.~\ref{sec:induction}, the} assessment of their general magnetic properties \WD{depends on Rm} and yields estimates for the leading contribution to electric currents $j$, amplitude of induced magnetic fields $B_\phi$ and the available Ohmic dissipation $P_{ohm}$. Tab.~\ref{tabinduction} shows these quantities \WD{that are discussed underneath for both planets in more detail. For both planets exist published, observation based atmospheric temperature profiles. In the case of KELT-9b, even day and nightside profiles were calculated. This more detailed analysis yields a deeper inspection of the thermal ionisation, electrical conductivity and subsequent induction process across various atmospheric layers.}

\begin{table}
\caption{Properties of the atmospheric induction based on the scaling laws and relations given in the text.}
\label{tabinduction}
\begin{tabular}{rccc}
\hline
 & & HD 209458b & KELT-9b \\
\hline
int. heat flux & $q_{int}$ [$\mathrm{W/m^2}$] & 5 .. $6.5 \times 10^3$  & 29 .. 553 \\
internal field & $B_{int}$ [mT] & 0.39 .. 4.3 & 0.72 .. 1.93  \\
electr. cond. & $\sigma_e$ [S/m] & $10^{-4}$ .. $10^{-2}$ & 1 .. 5  \\
length scale & $d$ [m] & $3 \times 10^5 $ & $2.7 \times 10^6$  \\
flow speed & $U$ [m/s] & $2 \times 10^3 $ & $1.3 \times 10^4$  \\
 $\mbox{Rm}$ &  & $\approx 1 $ & $4 \times 10^4$  \\
induced field & $B_\phi$ [mT] & $0.3 $ & 40 .. 400  \\
local current & j  [$\mathrm{A/m^2}$] & $0.8 \times 10^{-3}$ & 0.01 .. 0.1  \\
Ohmic heat & $P_{ohm}$  [$\mathrm{W}$] & $1.35 \times 10^{20} $ & $8.15 \times 10^{19}$ .. $5.34 \times 10^{21}$  \\
\hline
\label{tab:MHD}
\end{tabular}
\end{table}

\subsection{HD 209458b}
For HD 209458b - a typical HJ -  the electrical conductivity reaches values between $10^{-4}$ and $10^{-2} \,\mathrm{S/m}$ as shown in Fig.~\ref{ID_transport}, blue curves. The thermal  ionisation at this temperature causes only a partial ionisation of the potassium atoms (compare also the ionisation coefficient, $\alpha$) and thus a strong temperature sensitivity.  Using $d = 3 \times 10^5\, \mathrm{m}$ in \WD{Eq.~\ref{Rm}} this leads to $\mbox{Rm}\le 1$, suggesting that the magnetic diffusion indeed limits the growth of the induced field such that steady solutions are possible at $B_\phi = \mbox{Rm}\, B_{int}$.

The available estimates for the internal field strength $B_{int}$ are based on indirect magnetospheric observations and amount to 0.05~mT, a field strength ten times weaker than Jupiter's \citep{Kislyakova2014}. On the other hand, the field strength estimate based on the thermal evolution \WD{(\Eqnref{eqthevo})} and the energy balance \WD{(\Eqnref{eqThorn})} suggest higher values of 0.4 and 4~mT, respectively. HD 209458b is an inflated planet thus matching the proposed peak efficiency of Ohmic heating \citep{Thorngren2018}. This favours the energy balance estimate. 

Using $\sigma_e = 10^{-3}\,\mathrm{S/m}$, the local currents reach strength of roughly $10^{-3}\,\mathrm{A/m^2}$ and according to \Eqnref{eq:Ohmic:lin} a total atmospheric Ohmic power of $1.35 \times 10^{20}\,\mathrm{W}$ for an internal field strength of 0.4~mT. Using the upper field estimate of $4\WD{\,\mathrm{mT}}$ based on the energy balance would yield a two order of magnitude larger Ohmic dissipation matching the bolometric luminosity. The study of \citet{Batygin2010} used a 1~mT internal field and $\sigma_e \approx 10^{-2}\,\mathrm{S/m}$ and showed that roughly 1\% of the available Ohmic power is sufficient to explain the radius inflation.

For the deeper Ohmic heating \citet{Batygin2010} considered the region between $100\,$bar 
and $3000\,$bar where the conductivity increases from $\sigma_e=10^{-2}\,$S/m 
to $\sigma_i=10\,$S/m. This value should be used in \WD{\Eqnref{eq:Ohmic}}. Even though only a small fraction of the total Ohmic power is necessary to be deployed below the radiative-convective boundary, the larger uncertainty on the internal field strength and the strong temperature dependence of $\sigma_e$ makes a proper quantification very challenging.

\begin{table}
\caption{KELT-9b properties. The \WD{Brunt-V\"ais\"al\"a} frequency is calculated assuming an isothermal atmosphere $N_T = g /\sqrt{T_{eq} c_p}$ .}
\label{tabKELT9b}
\begin{tabular}{lccc}
\hline
name & variable &value & reference\\
\hline
planetary radius &$R_p$  & $1.354 \times` 10^8\,\mathrm{m}$ & \citet{Borsa2019} \\
planetary mass &$M_p$  & $5.467 \times 10^{27}\,\mathrm{kg}$ & \citet{Borsa2019} \\
Eq. temperature &$T_{eq}$ & $3921\,\mathrm{K}$ & \citet{Borsa2019} \\
surface gravity & $g$ & $20.8\,\mathrm{m/s^2}$ & \citet{Borsa2019}
\\
rotation frequency & $\Omega$ & 4.91 $\times 10^{-5}\,\mathrm{1/s}$ &\citet{Gaudi2017}\\
stratification & $N_T/\Omega$ &61.8 & \\
wind speed & U & $1.3 \times 10^4\,\mathrm{m/s}$ & \citet{Fossati2020} \\
density at 0.1 bar & $\rho$ & $7.16 \times  10^{-4}\,\mathrm{kg/m^3}$ & \WD{Fig.~2} \\
specific heat & $c_p$ & $1.2 \times 10^4\,\mathrm{J/kg\, K}$ & \\
radial scale height & $d_U$ & $2.7 \times 10^6\,\mathrm{m}$ &  2 \% radius\\

\hline
\end{tabular}
\end{table}

\subsection{KELT-9b}
As an \WD{high-temperature} counter example, we investigate the atmospheric structure and the induction process of KELT-9b. Planetary parameters from observations are summarised in tab.~\ref{tabKELT9b}, the derived quantities of the interior and atmospheric magnetic properties are given in tab.~\ref{tabinduction} and compared to HD 209458b.

\subsubsection*{ionisation degree and the electrical conductivity} \label{elec_trans_sec} 

\begin{figure}
\centering   
\includegraphics[width=8.5cm]{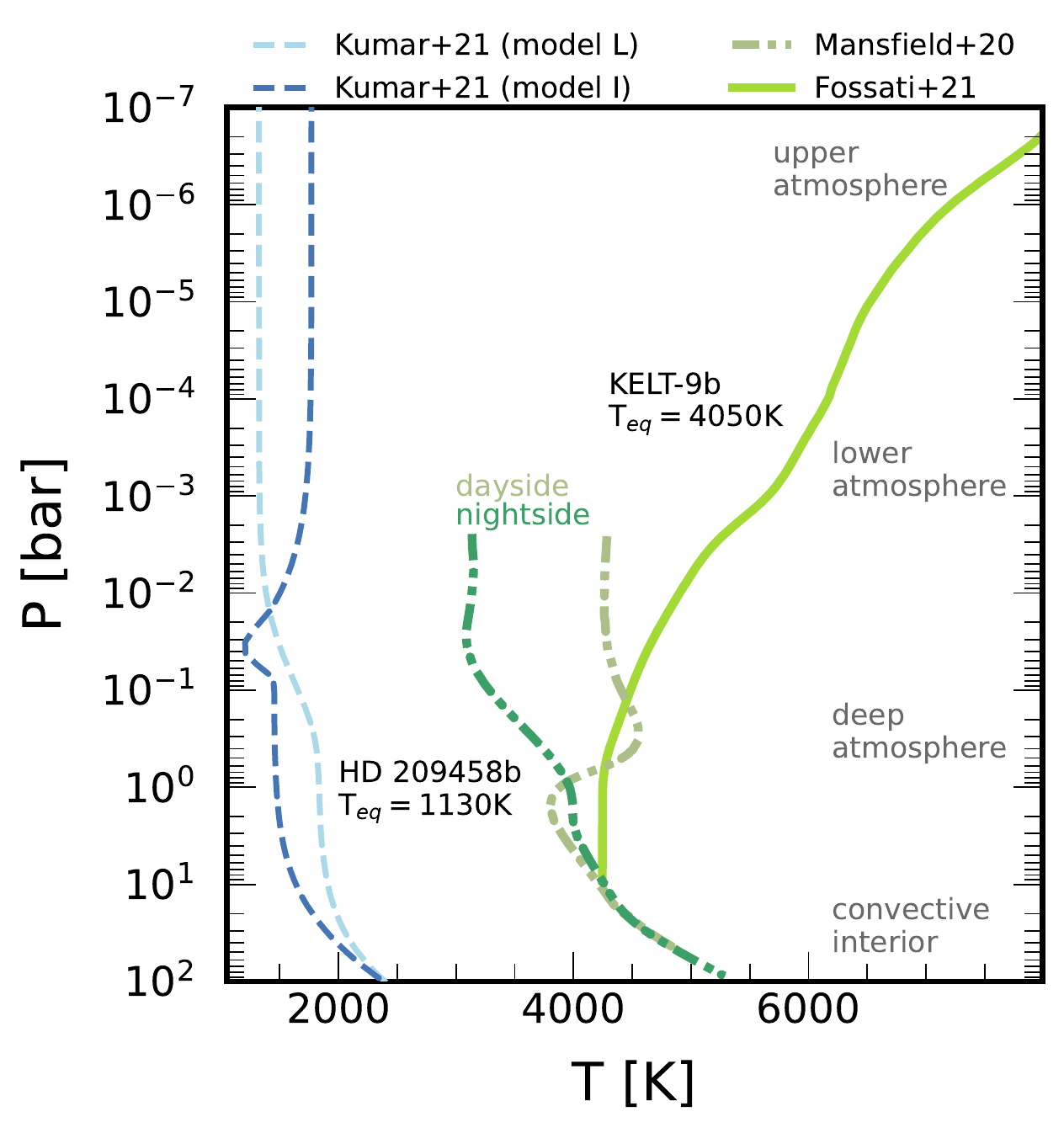}
\caption{Pressure-temperature profiles of the atmosphere of the ultra-hot Jupiter KELT-9b as used in our work to illustrate the dependence of the transport properties on the atmospheric conditions. The \WD{dayside} profile of \citet{Fossati2021} is shown as a solid line. The \WD{dayside and nightside} profiles by \citet{Mansfield2020} are displayed as dash-dotted lines. Additionally, the atmospheric profiles used in our previous work on the hot Jupiter HD~209458b~\citep{Kumar2021} are shown in \WD{blue}.}
\label{P_T_profile}	                           
\end{figure}

First, we calculate the ionisation degree $\alpha$ and the electrical conductivity $\sigma_e$ for $P$-$T$ conditions as assumed to be present in the atmosphere of KELT-9b. The $P$-$T$ profiles used for the calculation of the transport properties are fundamental input parameters to our induction model. The atmospheric conditions of KELT-9b have been extensively studied, e.g., to investigate the source of the high upper atmospheric temperatures indicated by observations of Balmer series and spectral lines of metal atoms~\cite{Hoeijmakers2018, Hoeijmakers2019, Wyttenbach2022}. \\
% what profiles are we using and why? introductory sentence...
We first summarise two studies whose outcomes we find adequate to use in our work, particularly the $P$-$T$ profiles presented in there.
% Mansfield:
\citet{Mansfield2020} present Spitzer phase curve observations of KELT-9b, deducing day- and nightside planetary temperatures of about $4600\,$K and $2600\,$K, respectively. They employ a global circulation model (GCM) to investigate the effect of additional heat transport mechanisms such as $H_2$ dissociation and recombination, which are included in the energy balance. The synthetic phase curves based on the GCM profiles yield better agreement with the observations than those without heat transport involving chemical reactions, but cannot reproduce the planetary temperatures derived from the phase curve observations. The corresponding $P$-$T$ profiles yield too small planetary temperatures with a temperature difference between the averaged day- and \WD{nightside} profiles of about $1000\,$K.
% Fossati:
\citet{Fossati2021} take a different approach to investigate the source of the upper atmospheric heating. They generate synthetic $P$-$T$ profiles and compare the ensuing synthetic transmission spectra with the observed H$\alpha$ and H$\beta$ line profiles. The inclusion of non-local thermal effects (NLTE) above $10^{-4}\,$bar in their models shows a strong influence on the atmospheric thermal balance, resulting in a very good agreement with the line profiles. Additionally, the model of \citet{Fossati2021} accounts for heat redistribution from day- to \WD{nightside} in such way that the \WD{dayside} temperature of $4600\,$K is met~\citep{Mansfield2020}.\\
% To conclude...
To illustrate the effect of different atmospheric conditions on the transport properties, particularly of varying temperature conditions on the day- and \WD{nightside}, we here use both the averaged day- and \WD{nightside} profiles from the global circulation model by \citet{Mansfield2020} (despite the fact that they yield too low temperatures as described above) as well as the \WD{dayside} profile by \citet{Fossati2021}. 
We display the three $P$-$T$ profiles for KELT-9b in Fig.~\ref{P_T_profile}. Note that the pressure ranges  of the profiles are different due to different scopes of the atmospheric studies. 
% Add HD 209458b to the figure
Additionally, we show the profiles from our previous work in which we investigated the transport properties in the atmosphere of the hot Jupiter HD~209458b~\citep{Kumar2021} with an equilibrium temperature of $1130\,$K for various $P$-$T$ profiles.

%%%%%%%%%%%%%%%%%%%%%%% Figure PT profiles

\begin{figure}
\centering   
\includegraphics[width = 8.5cm]{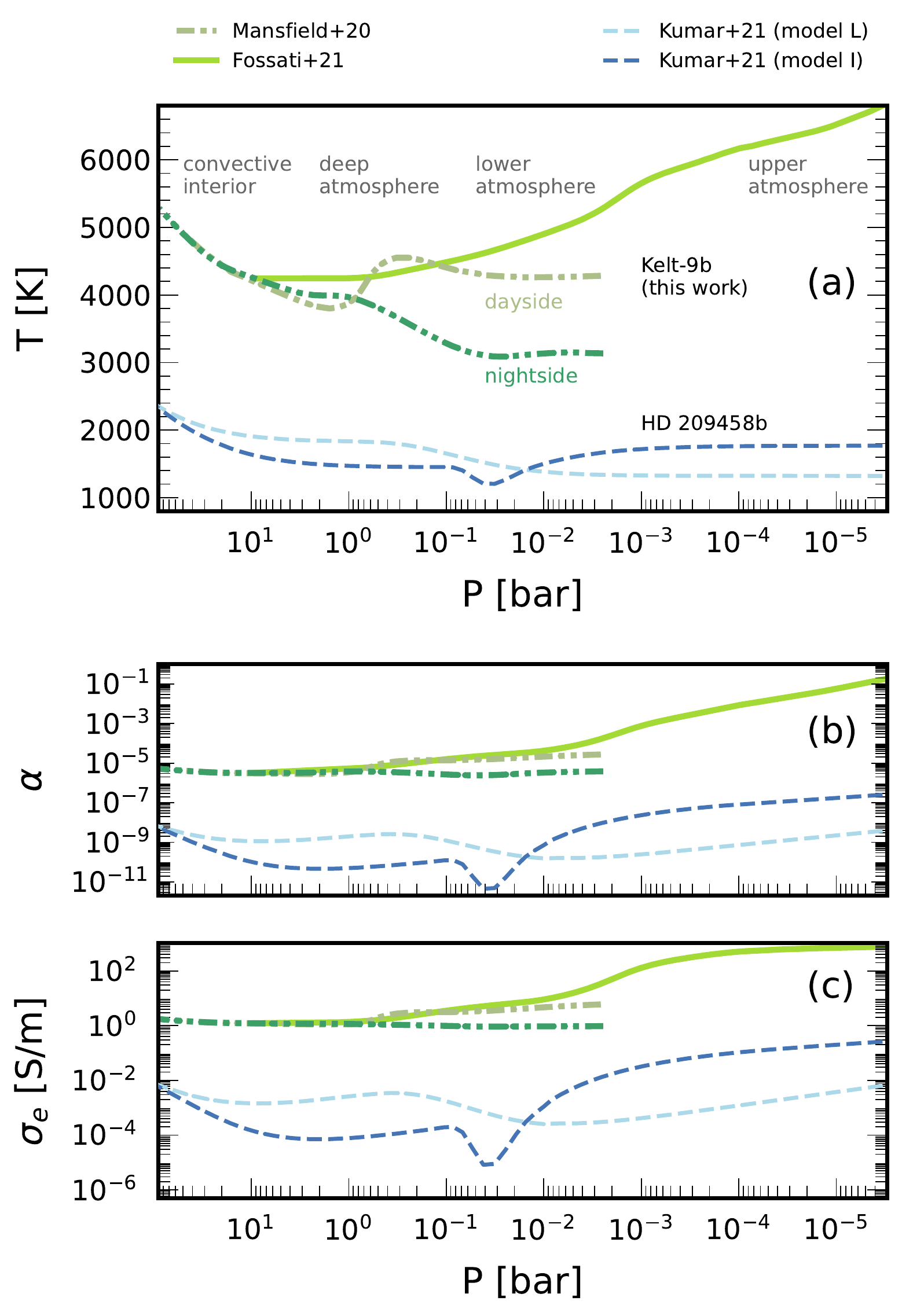}
\caption{Atmospheric temperature $T$, ionisation degree $\alpha$, and electrical conductivity $\sigma_e$ along pressure $p$ for KELT-9b. For comparison, two atmospheric profiles (L and I) of HD~209458b (dashed line) are also added in blue \citep{Kumar2021}.}
\label{ID_transport}	                           
\end{figure}

% display the results
Our results for the ionisation degree and conductivity along the $P$-$T$ profiles are shown in Fig.~\ref{ID_transport}. We also compare with results from our previous work for HD~209458b~\citep{Kumar2021}. In general, ionisation degree (Fig.~\ref{ID_transport} (b)) and electrical conductivity (Fig.~\ref{ID_transport} (c)) are closely related and follow a very similar behaviour as the pressure decreases.  

% results
% results Fossati+
The \WD{dayside} $P$-$T$ profile of KELT-9b shows a temperature inversion, i.e., the temperature increases with decreasing pressure in the outer atmosphere. The increasing temperature and decreasing density leads to a growing degree of thermal ionisation. Since the ionisation degree increases and the electrons scatter less frequently at lower density, the electrical conductivity $\sigma_e$ increases toward the outer atmosphere. In the isothermal region ($0.5-10$~bar), the conductivity remains nearly constant. 

% results Mansfield (day and nightside profiles)
The \WD{dayside-to-nightside} variation in temperature can be significant in ultra-hot Jupiters. Thus, the amount of thermally ionised constituents may also differ on each side. For that reason, we calculate the ionisation degree and electrical conductivity for both day- and \WD{nightside}. For the profiles of \citet{Mansfield2020}, the dayside temperature in the lower atmosphere is $\approx 1600$-$2000\,$K hotter than the \WD{nightside}, followed by an inversion in the deep atmosphere. The ionisation degree (Fig.~\ref{ID_transport} (b)) is decreasing in the isothermal region ($0.002$-$0.02$~bar) due to an increase in the pressure which is followed by a sharp decrease at $0.3$~bar. This sharp decrease is a consequence of the decrease of temperature in the inversion layer. The electrical conductivity is following the ionisation degree profile. 

Similarly to the \WD{dayside}, the ionisation degree on the \WD{nightside} is decreasing in the isothermal region ($0.001-0.01$~bar), which is followed by an increase due to the rising temperature in the outer atmosphere. The electrical conductivity is following a very similar behaviour as the ionisation degree. The $\sigma_e$ values in the lower atmosphere are 5-6 times lower than the dayside values, and in the deep atmosphere they are similar as on the dayside. 

% comparison to HD 209458b
In comparison with the colder hot Jupiter HD~209458b, 
$\alpha$ and $\sigma_e$ have higher values because of the higher temperatures for KELT-9b (Fig.~\ref{P_T_profile}, green profiles). However as the rather constant ionisation degree suggests, the temperature typical KELT-9b, even on the nightside are high enough to sustain a plasma in which the bulk of the alkali metals are singly ionised. In comparison to colder HJs, the electrical conductivity $\sigma_e$ does not show the strong temperature dependence characteristic for ionisation processes. 

The absolute values of electrical conductivity in KELT-9b are between two and four orders of magnitude higher than in HD~209458b. For a detailed discussion of the HD~209458b profiles, see \citet{Kumar2021}. 

Note that we have restricted our calculation of ionisation and electrical conductivity to as low as $10^{-6}$ bar pressure because at further lower pressure (outer atmosphere) non-equilibrium processes and magnetic field will impact these properties. This is a subject  left for future work.

%%%%%%%%%%%%% figure transport coefficients

\label{ind_mag_sec} 

\subsubsection*{Internal heat flux and magnetic field strength}

The strength of the internal, dynamo-generated magnetic field for KELT-9b is a function of the available buoyancy power. The associated internal heat flux can be based on the thermal evolution (\Eqnref{eqthevo}). KELT-9b is young (300 Myrs \citep{Borsa2019}), heavy and large. Thus following this scaling relation, the intrinsic heat flux is  $q_{int}=550\,\mathrm{W/m^2}$. On the other hand, the relation from \citet{Thorngren2018} or \Eqnref{eqThorn} yields a smaller heat flux of $q_{int} = 29\,\mathrm{W/m^2}$. The internal, dynamo generated magnetic field based on \Eqnref{eqBint} 
\WD{reaches between} 1 and 2~mT.

\subsubsection*{Atmospheric induction}
Fig.~2 illustrates that the electrical conductivity $\sigma_e$ in the atmosphere of KELT-9b (green profiles) is up to four order of magnitude higher than in HD 209458b. The profiles also show that the atmosphere of KELT-9b is to first order isothermal in the dynamically relevant pressure range between 0.01 and 1 bar. 
Electrical conductivities are of the order $1\,$S/m on the \WD{nightside} and can become one order of magnitude larger on the \WD{dayside}. Given the quite strong day-to-\WD{nightside} temperature contrast of 1500\,K, the associated contrast in $\sigma_e$ is rather small. This once more shows that the almost all alkali metals are ionised at such temperatures and the depth and lateral gradients remain rather small. 

The length scale $d$ entering the magnetic Reynolds number 
will thus be determined by the induction process itself. Whereas simulations suggest rather broad zonal flows \cite{Showman2002}, they might be more constrained in radius. As they are driven by irradiation gradients, the winds will not reach deeper than 1\,bar, where the optical depth in infrared wavelength range reaches unity. We therefore estimate a radial wind scale height of one pressure scale \WD{height}, i.e. $d_U = 0.02~R \approx 2700\,\mathrm{km}$. 
Assuming furthermore wind velocities of about $13\,$km/s 
\citep{Fossati2021} then yields magnetic Reynolds numbers
of about $4.5 \times 10^4$ for KELT-9b. Under these conditions, the time variability of the induced field remains a key contribution and the induction is strongly non-linear.

Thus the induction of atmospheric magnetic fields will be runaway process that \WD{cannot} be stopped by diffusion. Firstly we consider, Alfv\'en waves as a possible scenario (see Fig.~\ref{figindHJ}, c). The maximum azimuthal field during cycle of the Alfv\'en wave is given by \Eqnref{eq:Bphi1}
\be
 \BAS_\phi = \left(\mu_0\rho\right)^{1/2} \, \UAS_\phi \approx 400\,\mathrm{mT}
\eep
The oscillation period is the inverse Alfv\'en frequency, $t_A \approx 150$ days.
This field amplitude will not be reached, since the shear is suppressed  with regard to the force balance (\Eqnref{eqforcebalance} and Fig.~\ref{figindHJ}, d) at an amplitude of
\begin{equation}
    B_\phi \approx (\bs{\rho} \mu_0)^{1/2} (d_U \Omega U)^{1/2} \approx  40 \,\textrm{mT} \ .
    \label{eqAlfven}\
\end{equation}

However, at a very similar field strength of $50\,$mT, and according to \Eqnref{eq:BphiTayler} the Tayler-instability will set in with the form of a (\WD{$m=1$})-kink instability (see Fig.~\ref{fig_conduc_overview}, e). The radial scale of the instability given by \Eqnref{eq:lrlh} \keltNb\ the 
radial scale is somewhat smaller than $0.02\,R$, 
a value very similar to our assumed zonal flow scale $d_U$. This instability grows until the saturation field strength is reached. Using our \keltNb\ values suggest a field strength (\Eqnref{eq:Binst}) of roughly $400\,$mT, about an order of magnitude larger than the critical strength.

For the larger field amplitude of 400~mT, the related local electrical current in this non-linear regime is given by \Eqnref{eq:Ohmic:nl}:
\begin{equation}
    j =  0.1 \, \mathrm{A/m^2} \ ,
\end{equation}
where we used $d_U = 2.7 \times 10^6$~m. This leads to a total Ohmic power of $5.34 \times 10^{21}$. Compared to the bolometric luminosity of 
\begin{equation}
L_ = 4 R_P^2 \sigma_{SB} T_{eq}^4  \approx 10^{24}\,W    
\end{equation}
this is a negligible fraction suggesting that other mechanisms, such as tidal heating, dissipate much more energy and are responsible for excessive luminosity. 

Using the electrical conductivity for KELT-9b, the minimal radial flow speeds allowing for dynamo action in the atmosphere independent of the internal field is
\begin{equation}
    u_r = \frac{1}{\mu_0 \sigma_e d_r} \approx 10\, \mathrm{m/s} \ ,
\end{equation}
a value that is matched in numerical simulations and observations for KELT-9b \citep{Komacek2019}. An alternative is the Tayler-Spruit dynamo, where the emerging instability replenishes the poloidal field component and thus maintains the dynamo, seem a possibility. Dynamos based on the horizontal variation of the electrical conductivity seems unlikely given the small horizontal variation indicated by Fig.~\ref{ID_transport}, bottom panel.

\section{Discussion}
\label{concl} 
The small semi-major axis of Hot Jupiters and the synchronous orbits cause high atmospheric temperatures, at which a sizeable electrical conductivity is generated from the thermal ionisation of metals.

Here we show, that as the temperature increases several metals ionise and contribute step-wise to an increase of electrical conductivity $\sigma_e$, which can amount $10^{-4}$~S/m at 1500~K (potassium) and 1~S/m at 3000~K (sodium and calcium). The absolute values depend on the abundance of the particular metal species. Iron and hydrogen start to play a role only at temperatures in excess of 4000~K. Together with the atmospheric winds, which are driven by irradiation gradients and have been measured to reach velocities of a few km per second \citep{Snellen2010, Fossati2021}, strong electromagnetic currents are to be expected.

Thus, electromagnetic effects, such as Ohmic dissipation \citep{Batygin2010}, magnetic drag \citep{Perna2010}, or weakening of the azimuthal flows via Lorentz forces \citep{Rogers2014} that tend to equilibrate the strong azimuthal irradiation \WD{contrasts,} have been suggested to explain the large radii, strong day-to-night side brightness contrasts or infrared phase shifts \citep{Menou2012, Rogers2017, Showman2015}. All of those require a reliable estimate of the electrical conductivity, atmospheric flows pattern and of the induction of electrical currents, magnetic fields and their dissipation. 

Our results \WD{suggest} that Hot Jupiters can be categorised in two distinct groups depending on their temperature. Colder planets, e.g HD 209458b, host a linear induction process in the irradiated \WD{atmosphere,} characterised by $\mbox{Rm} = \sigma_e \mu_0 d U<1$, where the induced field is limited by magnetic diffusion \citep{Batygin2010,Kumar2021}. Here the electrical conductivity is exclusively due to the (partial) ionisation of potassium. 
 This approximation is valid only for Hot \WD{Jupiters} with an equilibrium temperature of $T_{eq} < 1500$~K. The induced atmospheric currents are then simply $j = \sigma_e U_\phi B_{int}$, i.e. a function of flow speed, the internal field strength and the electrical conductivity. The Ohmic power scales with $\sigma$ and the square of flow speeds and internal field strength.

In hotter planets with larger electrical conductivity the induction of atmospheric magnetic fields is very rapid ($\mbox{Rm}>>1$) and can only be saturated by non-linear effects, e.g. back-reaction of the Lorentz forces onto the flow or the emergence of magnetic instabilities. This requires new estimates for the 
electromagnetic induction effects. As an example, we have calculated ionisation degree and electrical conductivity in the atmospheric plasma of the ultra-hot Jupiter KELT-9b being the hottest Hot Jupiter observed so far. The day and night side temperature in the relevant part of the atmosphere around 0.001 and 1~bar, \WD{reach} 4600~K and 3000~K, respectively. At these temperatures, all sodium and calcium atoms are \WD{singly} ionised, whereas the ionisation of iron only starts to contributes on the \WD{dayside}. The electrical conductivity,  $\sigma_e$ reaches values of roughly $1\,\mathrm{S/m}$. Consequently, $\sigma_e$ is rather constant in the relevant $T$ range. Even for such day-to-\WD{nightside} temperature contrast the difference in $\sigma_e$ is less than on order of magnitude. This is in strong contrast to colder HJs, where even small temperature variations will lead to large horizontal and radial variations in $\sigma_e$ .

In addition, the atmospheric winds, driven by the 
irradiation gradients, are strong and thus lead to
high magnetic Reynolds number. For KELT-9b, we estimate
$\Rm=4\times 10^{4}$ and thus the induced magnetic field  will quickly outgrow the internal field. Radial field lines representing the internal field will be wound-up around the planet by the atmospheric shear. Thus the induced field $\BAS_\phi$ is predominantly axisymmetric and azimuthal. This process shares strong similarities with the solar tachocline, where the radial field is wound-up. We therefore relied on other fundamental theoretical considerations from planetary and stellar dynamo physics. 

We have discussed different \WD{mechanisms} that would limit this 
process and determine the saturated field strength of $\BAS_\phi$. 
The respective estimates suggest atmospheric, horizontal field strengths between $40\,$mT and
$400\,$mT. The larger values are based on the saturation field strength of the Tayler-instability. This is still an active area of research and should be seen with caution. 

The smaller \WD{estimate,} derived from the leading order force balance between the Coriolis and the Lorentz force, is more conservative and has thus \WD{higher
credibility}. At $40\,$mT the azimuthal field  
would be two orders of magnitude 
larger than Jupiter's observed field and also significantly 
larger \WD{than the} interior field of \keltNb. 
For all estimates of the induced field, the strength is independent of the internal field strength and the electrical conductivity, but scales for example with the rotation rate, the wind speed or atmospheric stratification. 

The axisymmetric, azimuthal field $\BAS_\phi$ 
stays inside the planet and cannot be observed, but it contributes to the internal dynamics via Lorentz forces and via Ohmic heating. Because of the efficient winding-up, the Lorentz forces will be so high that they play a substantial role in the atmospheric dynamics  

The associated electrical currents are calculated via $j = B_\phi / \mu_0  d_U $ and reach local values between 0.01 and 0.1~$\mathrm{A/m^2}$. For the entire layer involved in the atmospheric induction, the Ohmic Power amounts to between $10^{20}\,$W and $5 \times 10^{21}\,$W. 
This is a small fraction of the overall bolometric luminosity of roughly $10^{24}\,$W. This indicates that other processes, such as tidal heating are of greater importance in maintaining the high luminosity of KELT-9b. 

The weakness of the available Ohmic power ($\propto j^2/\sigma_e$) is a natural consequence of the non-linear induction process. I.e. the fact, that the induced field is independent of the electrical conductivity, thus the Ohmic power decreases with increasing $\sigma_e$.

Many Hot Jupiters are inflated. The degree of inflation seems to 
first increase with planetary equilibrium temperature $T_{eq}$ but decreases again beyond a maximum at about $1500\,$K 
\citep{Thorngren2018}. 
Several authors tried to explain the behavior with the fact that 
Lorentz forces slow down the atmospheric winds for higher $T_{eq}$ values \citep{Menou2012,Rogers2014}. 
Our analysis suggests a simpler explanation: 
The increased conductivity of the very hot planets 
makes deep Ohmic heating too inefficient.

We have only briefly touched on the possibility that additional 
overturning or radial motions in the stably stratified atmosphere could play in 
the induction process. These motions are bound to be slower than
the zonal winds but could nevertheless play an important role.
They will convert azimuthal field into smaller scale 
radial field, thereby limit the winding-up process but ultimately allow for an independent atmospheric dynamo. Our estimates for KELT-9b strongly \WD{suggests,} that radial flows of the order of a few tens of meters per second \WD{suffice} to turn the induction process into a self-sustained dynamo.

This bears the question whether both dynamos could be 
considered separately, as we have done here, or would 
influence each other. It has been suggested that 
a negative feedback between an internal and an external
dynamo could explain the weakness of Mercury's magnetic field 
\citep{Vilim2010,Heyner2011}.
In the model studied by \citet{Heyner2011} the addition of
an external magnetospheric dynamo quenched the overall
field strength by nearly three orders of magnitude. 
Such a coupled dynamo would prevent us from ever detecting
the magnetic field of an \UHJ.
Numerical dynamo simulations are required to 
explore these different options and to verify our 
theoretical predictions in the future.

\section*{acknowledgments}
We thank D. Shulyak for providing $P$-$T$ profile data of the atmosphere of KELT-9b. This work was supported by the Deutsche Forschungsgemeinschaft (DFG) within the Priority Program 
SPP~1992 ``The Diversity of Exoplanets" and the Research Unit FOR~2440 ``Matter under Planetary Interior Conditions''. This work was partially supported by the Center for Advanced Systems Understanding (CASUS) which is financed by Germany’s Federal Ministry of Education and Research (BMBF) and by the Saxon state government out of the state budget approved by the Saxon State Parliament.

\section*{appendix}
\subsection{Tayler-instability}
The mathematical description of this instability is based on the works of \citet{Spruit1999, Spruit2002} and has been derived in the context of the solar dynamo, where a \WD{stably} stratified zone of strong shear (tachocline) relentlessly creates azimuthal field from radial field lines. This winding up of the field was suggested to be stopped by the Tayler-instability.

The kinetic energy for this instability is 
provided by the restoring magnetic  force of the 
wound-up field $\B_\phi$. In analogy to the Alfv\'en
waves discussed above we conclude that the 
(maximum) kinetic energy is $1/2 \omega_A^2 \xi^2$,
where $\xi$ is a displacement perpendicular to the loop of induced azimuthal field lines. 
The stable stratification provides an obstacle which
requires the energy $1/2 N^2\xi_r^2$, where 
$\xi_r^2=\xi^2-\xi_h^2$ is the squared displacement 
in the direction of stratification and $\xi_h^2$ its
horizontal counterpart. The condition that the provided
kinetic energy should exceed the one required to work
against the stable stratification yields
$d_U< (\omega_A/N) d_h$ where we have assumed that the displacement reflects \WD{the}
radial scale $d_r$ and horizontal scale $d_h$ so that 
$d_r^2/d_h^2=\xi_r^2/\xi_h^2$.
Using $R_P$ as an upper bound
for $d_h$ then leads to the first condition for the 
instability that relates field strength and radial scale:
\bel{eq:lrlh}
   \frac{d_r}{R} < \frac{\omega_A}{N}
\eep

The second condition is that the growth rate 
of the instability must exceed the dissipation rate, that is \WD{$\eta / d_r^2 $}. The growth rate in the presence of a strong  Coriolis force is \WD{$ \omega_A^2 / \Omega$} \citep{Pitts1985}. 
Using \Eqnref{eq:lrlh}, 
the second condition provides a constraint
for the Alfv\'en velocity:
\bel{eq:dissgrowth}
    \frac{\omega_A}{\Omega} >  \left(\frac{N}{\Omega}\right)^{1/2} \left(\frac{\eta}{R^2\Omega}\right)^{1/4}
\eep
This translates into a critical field strength for
the onset of the Tayler-instability: 
\bel{eq:BphiTayler:a}
  \B_{\phi} = \Omega R \left(\mu_0\rho\right)^{1/2}\;
   \left(\frac{N}{\Omega}\right)^{1/2} \left(\frac{\eta}    
   {R^2\Omega}\right)^{1/4}
\eep

The magnetic continuity condition $\nabla\cdot\Bv=0$ 
relates instability components $\bm{B}^\prime$ and length scales:
\bel{eq:Binstscale}
  \frac{B_r^\prime}{B_\phi^\prime} \approx \frac{d_r}{R}
\eep
The primed field quantities denote the magnetic field
components of the Tayler-instability with the most
unstable azimuthal wave number $m=1$. 

The growth of the instability will stop once dissipation balances the rate at which the field is amplified. 
The amplification rate is given by \Eqnref{eq:Ind0}, 
\bel{eq:energyrate}
   \frac{\UAS}{d_r}\frac{\B_r^\prime}{\B_\phi^\prime}\approx 
   \frac{\UAS}{d_r}\frac{d_U}{R}
\eec
where we have used \Eqnref{eq:Binstscale}.
To quantify the effective dissipation rate in the presence of 
the instability, \citet{Spruit2002} assumes that it 
remains close to the growth rate of the instability: 
\bel{eq:dissrate2}
      \frac{\eta}{d_r^2}\approx \frac{\omega_A^2}{\Omega}
\eep

Combining  eqs.~\ref{eq:energyrate} and \ref{eq:dissrate2} 
and using condition \Eqnref{eq:lrlh} with the 
largest possible scale $d_r=\omega_A/ (N R_P)$ to
minimise dissipation yields:
\be
 \frac{\omega_A}{\Omega} = \frac{\UAS}{d_r\;N}
\eep
This translated into the field strength 
\bel{eq:Binst:a}
   \B_\phi = \left(\mu_0 \rho\right)^{1/2}\;
                \frac{\Omega}{N}\frac{R\,\UAS}{d_r}
\eep
While \Eqnref{eq:BphiTayler:a} estimates the field strength
at which the instability would set in, \Eqnref{eq:Binst:a}
estimates the field strength where it would saturate. 

The different assumptions made by \citet{Spruit2002}
have been criticised, in particular the way the 
saturation field strength \Eqnref{eq:Binst} has been derived 
(see for example \citet{Zahn2007} and \citet{Fuller2019}). 
So far, there seems to be no consensus. 
Note that estimate (\ref{eq:Binst}) can become smaller than
\Eqnref{eq:BphiTayler} for large values of $N/\Omega$. 
It seems likely that \Eqnref{eq:Binst} actually describes 
the field strength of the non-axisymmetric instability 
but unfortunately \citet{Spruit2002} is not clear about this. 

\section*{Data availability}
The data underlying this article will be shared on reasonable request to the corresponding author.

\bibliographystyle{mnras}
\bibliography{UHJ_bib}
\end{document}